\documentclass[sigplan,dvipsnames,screen,10pt]{acmart}
\usepackage{balance}
\usepackage[capitalize, noabbrev]{cleveref}
\usepackage{multicol}
\usepackage{caption}
\newcommand{\hs}{\lstinline}

\usepackage{listings}
\definecolor{darkblue}{rgb}{0,0,0.5}
\definecolor{darkgreen}{rgb}{0,0.3,0}
\definecolor{darkpink}{rgb}{0.4,0,0.3}
\definecolor{graygreen}{rgb}{0.3,0.5,0.3}
\definecolor{grayblue}{rgb}{0.2,0.2,0.6}
\definecolor{grayred}{rgb}{0.5,0.2,0.2}

\newcommand{\doublecolon}{$:\hspace{-0.4ex}:$}
\newcommand{\bindsymb}{\gg\hspace{-0.8ex}=}

\newenvironment{lfigure*}
  {\begin{figure*}
   \begin{flushleft}}
  {\end{flushleft}
   \end{figure*}}

\makeatletter
\lst@AddToHook{OnEmptyLine}{\vspace{-0.5\baselineskip}}
\makeatother

\AddToHook{cmd/appendix/before}{\crefalias{section}{appendix}}

    \def\noeditingmarks{}
  \usepackage{xargs}
  \usepackage[colorinlistoftodos,prependcaption,textsize=tiny]{todonotes}
  \ifx\noeditingmarks\undefined
      \newcommandx{\unsure}[2][1=]{\todo[linecolor=orange,backgroundcolor=orange!25,bordercolor=orange,#1]{#2}}
      \newcommandx{\info}[2][1=]{\todo[linecolor=green,backgroundcolor=green!25,bordercolor=green,#1]{#2}}
      \newcommandx{\change}[2][1=]{\todo[linecolor=blue,backgroundcolor=blue!25,bordercolor=blue,#1]{#2}}
      \newcommandx{\inconsistent}[2][1=]{\todo[linecolor=red,backgroundcolor=red!25,bordercolor=red,#1]{#2}}
      \newcommandx{\critical}[2][1=]{\todo[linecolor=purple,backgroundcolor=purple!25,bordercolor=purple,#1]{#2}}
      \newcommand{\improvement}[1]{\todo[linecolor=pink,backgroundcolor=pink!25,bordercolor=pink]{#1}}
      \newcommandx{\resolved}[2][1=]{\todo[linecolor=OliveGreen,backgroundcolor=OliveGreen!25,bordercolor=OliveGreen,#1]{#2}} % use this to mark a resolved question
    \else
      \renewcommand{\todo}{}
      \newcommandx{\unsure}[2][1=]{{}}
      \newcommandx{\info}[2][1=]{{}}
      \newcommandx{\change}[2][1=]{{}}
      \newcommandx{\inconsistent}[2][1=]{{}}
      \newcommandx{\critical}[2]{{}}
      \newcommand{\improvement}[1]{{}}
      \newcommandx{\resolved}[2][1=]{{}}

  \fi

\setcopyright{acmlicensed}
\acmDOI{10.1145/3759427.3760381}
\acmYear{2025}
\copyrightyear{2025}
\acmISBN{979-8-4007-2150-2/25/10}
\acmConference[OLIVIERFEST '25]{Proceedings of the Workshop Dedicated to Olivier Danvy on the Occasion of His 64th Birthday}{October 12--18, 2025}{Singapore, Singapore}
\acmBooktitle{Proceedings of the Workshop Dedicated to Olivier Danvy on the Occasion of His 64th Birthday (OLIVIERFEST '25), October 12--18, 2025, Singapore, Singapore}
\received{2025-06-09}
\received[accepted]{2025-07-31}

\begin{document}

\title{Invertible Syntax without the Tuples (Functional Pearl)}

\author{Mathieu Boespflug}
\orcid{0009-0005-3123-914X}
\affiliation{%
  \institution{Tweag}
  \city{Paris}
  \country{France}
}
\email{m@tweag.io}

\author{Arnaud Spiwack}
\orcid{0000-0002-5985-2086}
\affiliation{%
  \institution{Tweag}
  \city{Paris}
  \country{France}
}
\email{arnaud.spiwack@tweag.io}
%% If needed
% \renewcommand{\shortauthors}{Boespflug \& Spiwack}

\begin{abstract}
  In the seminal paper \emph{Functional unparsing}, Olivier Danvy used
  continuation passing to reanalyse printf-like format strings as
  combinators. In the intervening decades, the conversation shifted
  towards a concurrent line of work --- applicative, monadic or
  arrow-based combinator libraries --- in an effort to find combinators
  for invertible syntax descriptions that
  simultaneously determine a parser as well
  as a printer, and with more expressive power, able to handle
  inductive structures such as lists and trees. Along the way,
  continuation passing got lost. This paper argues that Danvy's
  insight remains as relevant to the general setting as it was to the
  restricted setting of his original paper. Like him, we present three
  solutions that exploit continuation-passing style as an alternative
  to both dependent types and monoidal aggregation via nested pairs,
  in our case to parse and print structured data with increasing
  expressive power.
\end{abstract}

%%
%% The code below is generated by the tool at http://dl.acm.org/ccs.cfm
%%


%%
%% The author(s) should pick words that accurately describe

\maketitle

\section{The Problem}
\label{sec:problem}

\citet{applicative} captured a common programming pattern as the
\hs{Applicative} class:
\begin{lstlisting}
class Functor f where
  (<$>)  :: (a -> b) -> f a -> f b

class Functor f => Applicative f where
  pure   :: a -> f a
  (<*>)  :: f (a -> b) -> f a -> f b
\end{lstlisting}
While the above is now standard in Haskell, \citeauthor{applicative}
note that the following alternative definition is equivalent:
\begin{lstlisting}
class Functor f => Monoidal f where
  unit   :: f ()
  mult   :: f a -> f b -> f (a, b)
\end{lstlisting}
One instance of programming with applicatives is parser combinator
libraries in the style of \citet{nhc-parsing}, also called
\emph{deserializers}. Conversely, combinators for pretty printing,
also called \emph{serializers}, are an instance of coapplicative
programming, characterized by reversing the arrow in the definition of the
\hs{Functor} class:
\begin{lstlisting}
class Contravariant f where
  (>$<)  :: (a -> b) -> f b -> f a

class Contravariant f => Comonoidal f where
  counit :: f ()
  comult :: f a -> f b -> f (a, b)
\end{lstlisting}
\textbf{What we want:} dual-use combinators, which we will call
\emph{format descriptors}, for both parsing and pretty printing.

We could characterize descriptors as profunctors (see
\emph{e.g.}\ \cite{monadic-profunctors}), which generalize both
functors and contravariants:
\begin{lstlisting}
class Profunctor p where
  dimap :: (b -> a) -> (c -> d)
        -> p a c -> p b d
\end{lstlisting}
with an instance like the following:
\begin{lstlisting}
data (pr :*: pa) a b = pr a :*: pa b

instance (Contravariant pr, Functor pa) =>
  Profunctor (pr :*: pa) where
    dimap f g (pr :*: pa) =
      (f >$< pr) :*: (g <$> pa)
\end{lstlisting}

\paragraph{Tuple Troubles}
We could, furthermore, define a new sequencing operator for
profunctors analogous to \hs{(<*>)} above:
\begin{lstlisting}
(***) :: (Comonoidal pr, Monoidal pa)
      => (pr :*: pa) a a'
      -> (pr :*: pa) b b'
      -> (pr :*: pa) (a, b) (a', b')
(pr :*: pa) *** (pr' :*: pa') =
    comult pr pr' :*: mult pa pa'
\end{lstlisting}
But herein lies \textbf{the problem}: chaining \hs{pr :*: pa}
combinators in a sequence leads to deeply nested pairs. Effectively,
it appears that we are forced to work in an awkward recursively
uncurried style, when currying is more natural, avoids noisy
wrapping/unwrapping of constructors, and potentially has better
performance.

\begin{example}
  Assume a record \hs{Rec} with three fields. Given (de)serializers
  \hs{bool}, \hs{char} and \hs{int}, a (de)serializer for \hs{Rec}
  might be of the form given in \hs{ex1} below:
\begin{lstlisting}
data Rec = Rec {a :: Bool, b :: Char, c :: Int}

unwrapRec Rec{a, b, c} = ((a, b), c)
wrapRec ((a, b), c) = Rec{a, b, c}

ex1 :: _ => (pr :*: pa) Rec Rec
ex1 = dimap unwrapRec wrapRec spec where
  spec = bool *** char *** int
\end{lstlisting}
\end{example}

The (un)wrapping functions (a pair of them per constructor of the data
type) can, and should, be derived generically. Nevertheless, such an
API requires frequent calls to adaptor functions to mediate between
elements of nested tuples and components of data types. Worse,
different programming patterns call for nesting to the left like
\hs{((a, b), c)} or to the right like \hs{(a, (b, c))}. But
associativity doesn't come free. We have to resolve any impedance
mismatch by hand with conversion functions. When an API has this
tendency, we shall say that the API exhibits \emph{tuple troubles}.

The uncurried style leaks into many derived combinators, including
higher-order combinators like folds and
generic combinators from standard abstractions like \hs{Applicative}
and \hs{Alternative}, which either need to be redefined based on
tupling, or duplicated \cite{partial-isomorphism}.

In short, we seek to dualize \citeauthor{nhc-parsing}'s immensely
popular style of parser combinators to parsing/printing pairs, but,
\begin{enumerate}
\item without compromise on the ergonomics, and
\item without compromise on expressivity, to support arbitrary
  structured data, including collections and trees (sums of products),
  not just records (products).
\end{enumerate}

In this paper, we present three solutions. The latter two are novel
and satisfy all of the above requirements, moreover in a setting
limited to Haskell'98 + rank-2 polymorphic types.

\section{Analysis}
\label{sec:analysis}

Consider the following variations of class methods introduced above:
\begin{lstlisting}
comult' :: Comonoidal f => (r -> (a, b))
        -> f a -> f b -> f r
comult' split f g = split >$< comult f g

mult'   :: Monoidal f => (a -> b -> r)
        -> f a -> f b -> f r
mult' combine f g =
  uncurry combine <$> mult f g
\end{lstlisting}
They reveal a deeply significant asymmetry between splitting comonoidally
and combining monoidally: \hs{combine} is a curried function of two
arguments, but flipping the arrow in the type of \hs{split} gives us
an uncurried function. This is the crux of the problem: there is no
notion of “curried” multiple-result function. This is why
\hs{Comonoidal} exhibits tuple troubles, while \hs{Applicative}
manages to avoid them. We need symmetry
to sequence covariant and contravariant functors simultaneously;
uncurrying functions everywhere is one way to recover that symmetry.
But there is another way.

The key insight in this paper is to invoke the folklore observation
that in continuation-passing style (CPS), multiple-argument functions and
multiple-result functions are completely symmetric, and
the types can be written with arrows alone. We need not pack multiple
results in a tuple, like in the \hs{Comonoidal} class.

Compare the types of 2-argument functions with 2-value functions in
CPS (with continuations passed first, not last):
\begin{lstlisting}
(s -> r) -> (a -> b -> r) -- two arguments
(a -> b -> r) -> (s -> r) -- two return values
\end{lstlisting}

\section{An Applicative Solution} %% Cassette
\label{sec:tr}

\citet{combinator-parsers} propose a surprisingly succinct variation
of the following interface for the primitives atop which most other
\citeauthor{nhc-parsing}-style parser combinators are derived:
\begin{lstlisting}
class _ p => Parsing p where
  satisfy :: (Char -> Bool) -> p Char
\end{lstlisting}
We can vary the expressive power of the combinators with different
choices to fill the constraint hole in the above.
\citet{combinator-parsers} use \hs{Alternative}. \citet{nhc-parsing}
uses \hs{Monad}. In this section, we will match the power of
\hs{Applicative}
i.e.\ \hs{Alternative} without choice. However, inspired by the
polyvariadic printing of \citet{functional-unparsing}, our setting
will be a more general notion of polyvariadic parsers, instead of the
unary parsers above.

\citeauthor{functional-unparsing} postulates an abstract syntax of
\emph{format descriptors}:
\begin{itemize}
\item \hs{lit} for declaring literal strings;
\item \hs{int} for specifying integers;
\item \hs{str} for specifying strings;
\item \hs{compose} to ``glue [descriptor] components together''.
\end{itemize}
This language of descriptors is used to specify how to pretty print (or
\emph{format}) input data, but can serve just as well to specify how
to parse it \cite{functional-ununparsing}. How would we do so, and can
we relate this to the \hs{Parsing} interface above?

\subsection{String Transformers}
\label{sec:string-transformers}

Recall the definition of a \hs{Category} as a structure with objects
and morphisms \hs{cat a b} from object \hs{a} to object \hs{b}, including
identity morphisms, all of which can be composed:
\begin{lstlisting}
class Category cat where
  id :: cat a a
  (.) :: cat b c -> cat a b -> cat a c
\end{lstlisting}
Now, consider the type of string transformers, and their type in CPS:
\begin{align}
  \textsf{String} &\to \textsf{String} \tag{0} \\
(\textsf{String} \to r) &\to (\textsf{String} \to r) \tag{1}
\end{align}
Functions always take a \emph{continuation} as their first argument.
The return type $r$ of continuations is called the \emph{answer type},
which in general is allowed to change from continuation to
continuation. Answer-type modification (ATM) is considered
a \emph{control effect}. So the most general type of string
transformers can be defined as below. String transformers form
a category under function composition:
\begin{lstlisting}
newtype Tr r r' =
  Tr ((String -> r) -> String -> r')

instance Category Tr where
  id = Tr id
  Tr f . Tr g = Tr (f . g)
\end{lstlisting}
Notice that parsing and pretty printing both fit within the more
general framework of CPS string transformers:
\begin{lstlisting}
Tr r (a -> r)                -- For printing
Tr (a -> r) r                -- For parsing
\end{lstlisting}
We can check this by expanding the definition of \hs{Tr}. A string
transformer for parsing provides the parsed value and the remaining
string to the continuation. For printing, a string transformer expects
an input string and a value to print, prints the value to another
string, and provides its continuation with the concatenation of these
two strings:
\begin{lstlisting}
(String -> r) -> String -> a -> r  -- print
(String -> a -> r) -> String -> r  -- parse
\end{lstlisting}
Notice that here string transformers for parsing are those for
printing with a flipped top-most arrow. The first type is an
instance of the type of (printing) format descriptors in
\cite{functional-unparsing}.
In printers, \hs{r} is typically polymorphic, such that with
polymorphic functions \hs{f :: Tr (a -> r) r} and \hs{g :: Tr (b -> r) r}, instantiating \hs{r} in the
type of \hs{f} gives, \hs{(f :: Tr (a -> b -> r) (b -> r))}, so that
the composition \hs{f . g} has type \hs{Tr (a -> b -> r) r}, again for an
arbitrary \hs{r}. Type instantiation, like this, is the essence of composition in \cite{functional-unparsing}.

The type for parsing is novel\footnote{Novel but flawed: if \hs{f :: Tr r (a -> r)} and \hs{g :: Tr r (b -> r)}, then \hs{f . g :: Tr r (b -> a -> r)}. We want \hs{a}, \hs{b} in the oppposite order.}. But indeed, there are multiple
ways to construct a type of parsing descriptors with \hs{Tr}. An alternative
is to instantiate the answer types for printing with flipped
polarities\footnote{Recall that a type occurrence is positive if it
lies to the right of an even number of arrows starting from the top
level, and negative if to the right of an odd number.}:
\begin{lstlisting}
Tr r (a -> r)                -- For printing
Tr (r -> t) ((a -> r) -> t)  -- For parsing
\end{lstlisting}
This type for parsing string transformers is an instance of the type
of (parsing) format descriptors proposed by
\citet{functional-ununparsing}.

Like for the type of printing descriptors, gluing parsing descriptors to form
new descriptors is composition of morphisms, \emph{i.e.}\ just function
composition. Therefore, a pair of printing/parsing string
transformers, which we'll call a \emph{cassette} (or \hs{K7} for short), is itself an
instance of \hs{Category}\footnote{We will see later why using lazy
irrefutable pattern matching is important.}:
\begin{lstlisting}
data K7 p a b = K7
  { sideA :: p a b,
    sideB :: forall t!.! p (a -> t) (b -> t) }

instance Category p => Category (K7 p) where
  id = K7 id id
  ~(K7 f f') . ~(K7 g g') =
      K7 (f . g) (f' . g')

type PP a = forall r!.! K7 Tr r (a -> r)
type PP0 = forall r!.! K7 Tr r r
\end{lstlisting}
Cassettes have a tape with a pair of tracks. The \hs{Category} instance tells us that the
tapes of two cassettes can be spliced to form a new
cassette.
Crucially, splicing is associative: \hs{a . (b . c)} and \hs{(a . b) . c} denote the same cassette.

A \hs{PP a} is simultaneously a descriptor for printing and for parsing values
of type \hs{a}. In the framework of string transformers, this is but
a special case of descriptors of any arity, so we introduce type
synonym \hs{PP0} for nullary descriptors.

\subsection{Primitive and Derived String Transformers}
\label{sec:prim-deriv-tr}

\hs{satisfy} is a primitive descriptor to print/parse a single
character satisfying the given predicate:
\begin{lstlisting}
satisfy :: (Char -> Bool) -> PP Char
satisfy p = K7 (Tr pr) (Tr pa) where
  pr k s c = k (s ++ [c])
  pa k s@(c:cs) u | p c = k cs (u c)
\end{lstlisting}
We now have enough to implement \hs{lit}, the descriptor that
prints/parses a literal string, as a derived combinator:
\begin{lstlisting}
lit :: String -> PP0
lit [] = id
lit (c:cs) = satisfy (== c) . lit cs
\end{lstlisting}
We can translate between curried and uncurried style with the
following primitive:
\begin{lstlisting}
pairL :: K7 Tr (a -> b -> r) ((a, b) -> r)
pairL = K7 (Tr (uncurry .)) (Tr (curry .))
\end{lstlisting}
Finally, it is useful to afford ourselves an equivalent to \hs{(<$>)}
in applicative-style parsing, to map over a descriptor. In our setting,
this is possible given any pair of morphisms that are inverses of each
other (i.e.\ an isomorphism), one for printing and one for parsing:
\begin{lstlisting}
isoL :: (s -> a) -> (a -> s)
     -> K7 Tr (a -> r) (s -> r)
isoL to from = K7 (Tr pr) (Tr pa) where
  pr k s t = k s (to t)
  pa k s u = k s (\x -> u (from x))
\end{lstlisting}

\begin{example}\label{ex:scanf-printf}
  Typed \verb|sprintf()| \cite{functional-unparsing,typed-functional-unparsing} and its inverse
  \verb|sscanf()| \cite{functional-ununparsing} are both implementable
  in our framework:
\begin{lstlisting}
sprintf :: K7 Tr String r -> r
sprintf (K7 (Tr pr) _) = pr id ""

sscanf :: K7 Tr r r' -> String -> r' -> r
sscanf (K7 _ (Tr pa)) s k = pa (const id) s k
\end{lstlisting}
  Given the following derived combinators,
\begin{lstlisting}
char :: PP Char
char = satisfy (const True)

digit :: PP Int
digit = conv . satisfy isDigit where
  conv = isoL (head . show) (\c -> read [c])
\end{lstlisting}
  and the following format descriptor,
\begin{lstlisting}
spec = digit . lit "-th character after " .
       char . lit " is " . char
\end{lstlisting}
  we can implement an example due to \citet{functional-ununparsing}:
\begin{lstlisting}
>>> sprintf spec 5 'a' 'f'
"5-th character after a is f"

>>> sscanf spec "5-th character after a is f" (,,)
(5, 'a', f')
\end{lstlisting}
The inferred type of \hs{spec} is \hs{K7 Tr r (Int -> Char -> r)}.
This is the type of a binary descriptor.
\end{example}
Where in the \hs{Parsing} interface, descriptors are applicative
combinators, which can be sequenced \hs{(<*>)} and mapped over \hs{(<$>)}, here descriptors are
modeled as morphisms, which compose \hs{(.)} and can be mapped over with \hs{isoL}.
The respective operators are in 1:1 correspondence.
\hs{pure} corresponds to mapping over \hs{id} with \hs{isoL}.
We have
the same expressive power, in a different formalism.
What we have gained is printing with the descriptors, not just parsing.

\section{An Alternative Solution} %% Cassette
\label{sec:c-tr}

The \hs{char} and \hs{digit} descriptors of Example~\ref{ex:scanf-printf} only match a single character
from the input. The full \hs{str} and \hs{int} descriptors of
\citeauthor{functional-unparsing} are, on the parsing side, examples of \emph{maximal
munch}, which means matching as long of a portion
of the input as possible. To derive them as iterating \hs{char} and
\hs{digit} requires a new primitive encoding the pattern, or in general, a notion of
choice, to specify when to continue munching or to stop.

Returning to the \hs{Parsing} interface, in this section we choose
\hs{Alternative} as the superclass, whose two methods serve to add
failure and choice to applicatives, respectively:
\begin{lstlisting}
class Applicative f => Alternative f where
  empty  :: f a
  (<|>)  :: f a -> f a -> f a
\end{lstlisting}
This interface has sufficient expressive power to
express context free grammars \cite{combinator-parsers}.

\subsection{Vertical Composition}

Our morphisms are not applicatives, but we can likewise add similar
structure:
\begin{lstlisting}
class Monoid a where
  mempty :: a
  (<>) :: a -> a -> a
\end{lstlisting}
If we could make string transformers an instance of \hs{Monoid}, we
would horizontally compose morphisms using \hs{(.)}, and vertically
compose using \hs{(<>)}. Both operators have units.

\begin{example}\label{ex:skip}
  We can optionally apply a nullary descriptor. If it fails, that's
  fine --- we'll just do nothing in that case:
\begin{lstlisting}
optional :: PP0 -> PP0
optional p = p <> id
\end{lstlisting}
\end{example}

Cassettes are monoids if the underlying category is itself a monoid:
\begin{lstlisting}
instance (forall r r'!.! Monoid (p r r')) =>
    Monoid (K7 p r r') where
  mempty = K7 mempty mempty
  K7 f f' <> K7 g g' = K7 (f <> g) (f' <> g')
\end{lstlisting}
So we need to adapt \hs{Tr} to make it a monoid. Parsers that
sometimes fail are often implemented the same way we model partiality:
with the \hs{Maybe} monoid. For example, if a parser fails, it does
not yield a value nor the rest of the string. It returns \hs{Nothing}.
We'll need partiality for printing as well, so that both parse
descriptors and descriptors can uniformly be composed vertically to
incrementally build a total descriptor describing how to both print and
parse.

\begin{example}\label{ex:true-false}
  Assume a variant \hs{lit' :: PP ()} equivalent to \hs{lit :: PP0}
  and a primitive \hs{prismL} to lift a prism \cite{profunctor-optics}
  to string transformers (more on prisms in \cref{sec:prisms}).
  We can construct a (total) descriptor for formatting boolean literals
  piecewise from (partial) descriptor components, one for each
  constructor of data type \hs{Bool}:
\begin{lstlisting}
_True :: Prism' Bool ()
_False :: Prism' Bool ()

true, false :: PP Bool
true = prismL _True --> lit' "T" where
false = prismL _False --> lit' "F" where

bool = true <> false
\end{lstlisting}
\end{example}

The problem is that there is no uniform way to introduce \hs{Maybe} in
the answer types of both sides of a cassette. On the printing side,
say the continuation has return type \hs{Maybe r}. Then we can no
longer compose two printing string transformers. For example,
\begin{lstlisting}
>>> let p = _ :: Tr (Maybe r) (a -> r')
>>> p . p
-- Type error: can't match a->r' with Maybe r.
\end{lstlisting}

So what would Olivier Danvy do? Perhaps observe that there is no
problem that cannot be solved with an extra pass of the CPS
transform\footnote{See \emph{e.g.}\ \cite[§3]{type-directed-pe} for
one instance of that.}! Well then, let's continue the series of string
transformer types one more level --- that corresponding to level 2 in
the CPS hierarchy \cite{abstracting-control}:
\begin{align}
\mbox{\hs{String}} &\to \mbox{\hs{String}} \tag{0} \\
\mbox{\hs{(String -> r)}} &\to \mbox{\hs{(String -> r)}} \tag{1} \\
\mbox{\hs{((String -> r) -> r)}} &\to \mbox{\hs{((String -> r) -> r)}} \tag{2}
\end{align}
Iterating the CPS transform one more time gives us access to a second
continuation. We will use this continuation as a failure continuation
\cite{nhc-parsing}. By convention, we will say \hs{k} for the success
continuation and \hs{k'} for the failure continuation.

The definition of \hs{Tr} now reads as follows, again allowing for
ATM:
\begin{lstlisting}
type C r = (String -> r) -> String -> r

newtype Tr r r' = Tr { unTr :: C r -> C r' }
\end{lstlisting}
As before, \hs{Tr} is an instance of \hs{Category}. The definitions of
these instances remain unchanged. But we can now add a new instance:
\begin{lstlisting}
instance Monoid (Tr r r') where
  mempty = Tr \_ k' s -> k' s
  Tr f <> Tr g =
    Tr \k k' s -> f k (\_ -> g k k' s) s
\end{lstlisting}
\hs{mempty} is the string transformer that throws away the success
continuation because it always fails. And where \hs{(.)} is
(horizontal) composition of success continuations, \hs{(<>)} is
(vertical) composition of failure continuations. If the first
transformer fails, the computation proceeds with the second
transformer, with the same input string and success continuation as
initially.

We need to adapt \hs{satisfy} to the new definition of \hs{Tr}:
\begin{lstlisting}
satisfy :: (Char -> Bool) -> PP Char
satisfy p = K7 (Tr pr) (Tr pa) where
  pr k k' s c
    | p c = k (\s -> k' s c) (s ++ [c])
    | otherwise = k' s c
  pa k k' (c:cs) u
    | p c = k (\cs _ -> k' cs u) cs (u c)
  pa _ k' s u = k' s u
\end{lstlisting}
Unlike in \cref{sec:tr}, \hs{satisfy} is now a total function.
Printing and parsing both only succeed if the supplied predicate
succeeds. Otherwise, we call the failure continuation \hs{k'}, instead
of leaving that case undefined.

\subsection{Leads}
\label{sec:leads}

Vertical composition opens up new possibilities, not just of failure.
Consider again the type of \hs{spec} in Example~\ref{ex:scanf-printf}:
\begin{lstlisting}
spec :: K7 Tr r r' !\textrm{where}! r' = Int -> Char -> r
\end{lstlisting}
We can view a cassette as an abstract machine with a tape of instructions and a
stack. The types \hs{r} and \hs{r'} respectively give us the precondition and postcondition of
the stack when the machine reads the tape in one direction and executes the sequence of instructions that if finds, represented by \hs{spec}. Of course, since this paper is about invertible syntax, the \hs{spec} can also be executed in the other direction, in which case the pre- and post- conditions are reversed.

In \cref{sec:stacked-monads}, we'll see how to define the generic
\hs{push} and \hs{pop} operations one would expect to manipulate the
stack. For now, consider that primitive descriptors like \hs{char}, \hs{digit} also affect the stack:
\begin{lstlisting}
id              :: K7 Tr r r
char . id       :: K7 Tr r (Char -> r)
int . char . id :: K7 Tr r (Int -> Char -> r)
\end{lstlisting}
We can read this sequence of primitive descriptors as instructions.
From right to left, each instruction pushes a new type on the stack.
In the opposite direction, from left to right, these instructions pop
off the stack.

In this light, we can view quantifying over all \hs{r} as a kind of
frame rule: a descriptor of type \hs{PP a} behaves the same no matter the
state of the stack.

Some format descriptors simultaneously push and pop from the stack.
Consider the
following string transformer \hs{consL}. It pushes the two types of the components of the \hs{(:)} constructor on the stack, after having popped the result type off the stack, or \emph{vice versa}:
\begin{lstlisting}
id         :: K7 Tr r r
consL . id :: K7 Tr (a -> [a] -> r) ([a] -> r)
\end{lstlisting}
On the parsing side, \hs{consL} constructs a list using \hs{(:)}. To do so, it pops the components off the stack and pushes the result on the stack.
On the printing side, time flows in reverse. \hs{consL} expects a list on the stack.
It pops the list to deconstruct
(i.e.\ pattern matches on) it, before pushing the components onto the
stack if the list is non-empty.

We call such string transformers \emph{leads}, by analogy to the
lead-ins and lead-outs at the beginning and at the end of the tape in
a cassette. In our setting, we have lead-ins on the printing side and
lead-outs on the parsing side. Lead-ins and lead-outs have types of
the following form, respectively:
\begin{lstlisting}
Tr (b!$_1$! -> !$\ldots$! -> b!$_n$! -> r)        (a -> r)
Tr((b!$_1$! -> !$\ldots$! -> b!$_n$! -> r) -> t) ((a -> r) -> t)
\end{lstlisting}
These types generalize the types of printing and parsing
string transformers from \cref{sec:string-transformers}, respectively.

Intuitively,
\begin{itemize}
\item a \emph{lead-in} deconstructs a value if possible and passes the
  components to its continuation, or fails otherwise;
\item a \emph{lead-out} constructs a value from components and passes
  that to its continuation. It never fails.
\end{itemize}
A pair of a lead-in and a lead-out is a lead. By convention, we name
leads with an \hs{*L} suffix. We can define a lead like \hs{consL} by
hand:
\begin{lstlisting}
consL = K7 (Tr leadin) (Tr leadout) where
  leadin k k' s xs@(x:xs') =
    k (\s _ _ -> k' s xs) s x xs'
  leadin _ k' s xs = k' s xs
  leadout k k' s u =
    k (\s _ -> k' s u) s (\x xs' -> u (x:xs'))
\end{lstlisting}
But more typically, we'll build leads from existing prisms, as in the
next section.

\subsection{Prisms}
\label{sec:prisms}

A \emph{prism} is an optic used to work with sum types
\cite{profunctor-optics}. Like a lens, it is a bidirectional
structure, relating the value of a data type to its components. It
offers a \emph{review} operation to inject components into a sum type,
and a \emph{preview} to project these components. Reviewing always
succeeds, whereas previewing can fail. Notice the similarity with
leads: a lead-in is a preview and a lead-out is a review. Therefore,
we can lift all prisms to lead string transformers:
\begin{lstlisting}
prismL :: Prism' s a -> K7 Tr (a -> r) (s -> r)
prismL l = K7 (Tr leadin) (Tr leadout) where
 leadin k k' s t = case preview l t of
   Nothing -> k' s t
   Just x -> k (\s _ -> k' s t) s x
 leadout k k' s u =
   k (\s _ -> k' s u) s (\x -> u (review l x))
\end{lstlisting}
This is convenient because prisms are well understood and well
supported. In practice, they are derived systematically, whether
through meta- or generic programming \cite{tyt,generic-lens}.

\begin{lfigure*}
\begin{minipage}{0.48\textwidth}
\begin{lstlisting}
type Idnt = String
data Term =
  Var Idnt | Abs Idnt Term | App Term Term

varL = prismL _Var
absL = prismL _Abs . pairL
appL = prismL _App . pairL
\end{lstlisting}
\end{minipage}
\begin{minipage}{0.51\textwidth}
\begin{lstlisting}
parens p = lit "(" . p . lit ")"
sepSpace = lit " "
idnt = consL !$\longrightarrow$! letter . many alphaNum

term :: PP Term
term = varL --> idnt <>
       absL --> lit "!$\lambda$!" . idnt . lit "!.!" . term <>
       appL --> parens (term . sepSpace . term)
\end{lstlisting}
\end{minipage}
\caption{The syntax of the pure $\lambda$-calculus.}
\label{fig:lc-grammar-cassette}
\end{lfigure*}

\subsection{Derived String Transformers}

We now have enough rope to implement a raft of new combinators. For
lack of space, we'll demonstrate the use of leads to implement just
two of them that are standard in \citeauthor{nhc-parsing}-style parser
combinator libraries:
\begin{lstlisting}
many :: PP a -> PP [a]
many p = some p <> nilL

some :: PP a -> PP [a]
some p = consL . p . many p
\end{lstlisting}
These higher-order descriptors
repeat the given descriptor, collecting the result at
each step. With these we can finally implement \hs{int}:
\begin{lstlisting}
int :: PP Int
int = isoL show read . some (satisfy isDigit)
\end{lstlisting}
Recursive combinators like these crucially depend on the composition
operator being lazy in both arguments, lest evaluation never
terminate. This explains the irrefutable patterns in
\cref{sec:string-transformers}.

\subsection{Case Study: The $\lambda$-Calculus}
\label{sec:case-lambda}

We defined \hs{sprintf} and \hs{sscanf} in
Example~\ref{ex:scanf-printf}. Analogues in the more general case,
where we handle sums in addition to products, are given below:
\begin{lstlisting}
pretty :: PP a -> a -> Maybe String
pretty (K7 (Tr f) _) =
  f (\_ -> Just) (\_ _ -> Nothing) ""

parse :: PP a -> String -> Maybe a
parse (K7 _ (Tr f')) s =
  f' (\_ _ x -> Just x) (\_ _ -> Nothing) s id
\end{lstlisting}
The initial success continuation of \hs{pretty} throws away the
failure continuation and returns just the result string, nothing
otherwise. For \hs{parse}, we throw away the failure continuation as
well as the unparsed rest of the string in case of success, nothing
otherwise.

We now have enough infrastructure in place to demonstrate the
generalization of \citeauthor{functional-unparsing}'s format descriptors
to context-free grammars. \cref{fig:lc-grammar-cassette} specifies the
syntax of a small programming language, the pure $\lambda$-calculus,
in the style of a grammar expressed in Backus-Naur form (BNF). We have one production, \hs{term}, with several alternatives. For
effect, we use \hs{(-->)}, a synonym for \hs{(.)}, to separate leads (which are pure string transformers) from the descriptor for components. The leads in this grammar are constructed from ambient
prisms derived automatically \cite{generic-lens}.

At the prompt, parsing the self-application term gives the result we
expect, and we can pretty print it back:
\begin{lstlisting}
>>> parse term "!$\color{grayred}\lambda$!x.(x x)"
Just (Abs "x" (App (Var "x") (Var "x")))

>>> (parse term "!$\color{grayred}\lambda$!x. (x x)") >>= pretty term
Just "!$\color{grayred}\lambda$!x.(x x)"
\end{lstlisting}

\section{A Monadic Solution} %% PUP
\label{sec:pup}

Expressive power is a limitation of the \hs{K7}-style format
descriptors. Parsers' control flow is static; they can't observe an
intermediate result to change how the rest of the input is parsed. In
effect, only context-free grammars can be parsed. While this is
a standard limitation\,--\,parser generators also exhibit static
control flow\,--\,we now turn to investigating how to parse languages
generated by context-sensitive grammars, \emph{e.g.}\ XML, in a single
pass.

In this section, we seek to match the expressive power of
\hs{Parsing} when we take \hs{Monad} as its superclass.
In doing so, we will have to forgo some of the
symmetry of cassettes. In exchange, we will build some reusable abstraction and
gain the ability to reuse existing parsing and pretty printing libraries, instead
of rolling our own.

\subsection{Stacked Monads}
\label{sec:stacked-monads}

As a starting point, let us turn to an established interface combining features
of categories (like in \cref{sec:tr,sec:c-tr}) and monads, namely \emph{indexed
monads} \cite{indexed-monads}. For the sake of completeness,
\cref{fig:indexed-functor-applicative} lists the derived combinators from the
\hs{IxMonad} class that we'll be using.
\begin{lfigure*}
\begin{minipage}{0.52\textwidth}
\begin{lstlisting}
(<$>) :: IxMonad m => (a -> b) -> m r r' a
      -> m r r' b
f <$> ma = ma >>= \a -> return (f a)

(<*>) :: IxMonad m => m r' r (a -> b) -> m r'' r' a
      -> m r'' r b
mf <*> ma = mf >>= \f -> ma >>= \a -> return (f a)
\end{lstlisting}
\end{minipage}
\begin{minipage}{0.47\textwidth}
\begin{lstlisting}
(<*) :: IxMonad m => m r' r a -> m r'' r' ()
     -> m r'' r a
ma <* mu = ((\a _ -> a) <$> ma) <*> mu

(*>) :: IxMonad m => m r' r () -> m r'' r' a
     -> m r'' r a
mu *> ma = ((\_ a -> a) <$> mu) <*> ma
\end{lstlisting}
\end{minipage}

  \caption{Functorial and applicative combinators for indexed monads}
  \label{fig:indexed-functor-applicative}
\end{lfigure*}

\begin{lstlisting}
class IxMonad m where
  return :: a -> m r r a
  (>>=) ::
    m r' r a -> (a -> m r'' r' b) -> m r'' r b
\end{lstlisting}
In \hs{m r' r a}, \hs{r} and \hs{r'} are the shapes of the stack before and
after printing, respectively, while the parser has \hs{a} as the type of its result,
instead of pushing its result to the stack like in \cref{sec:tr,sec:c-tr}. In
effect, we're decoupling the covariant and the contravariant arguments, in a
similar style to the profunctors of \cref{sec:problem}.

A pair of categories form a category.
A cassette is then a pair of two categories: print and parse format descriptors.
Similarly, a pair of indexed monads is an indexed monad, the detailed
implementation of which can be found, together with all the code supporting this
section, in \cref{sec:full-stacked-monad-linear}.

So we're looking for an indexed monad for printing, and another for parsing. As
remarked in \cite{monadic-profunctors}, the parsing case is easily solved:
take any regular parser monad \hs{m}, and lift it to an indexed monad by
ignoring the stack:

\begin{lstlisting}
newtype Fwd m r r' a = Fwd { unFwd :: m a }

instance Monad m => IxMonad (Fwd m) where
  return x = Fwd (return x)
  (Fwd a) >>= f = Fwd (a >>= (unFwd . f))
\end{lstlisting}

We shall therefore turn, in the rest of this section, our attention to the much less
obvious case of printers. Indeed, in order to make an (indexed) monad, printers
can't ignore the return type, since monad lets you branch on the return value.
The solution to this conundrum can be found, again, in
\cite{monadic-profunctors}: make printers return the value they print.

\paragraph{Continuations as an Indexed Monad}
To implement printers, we shall turn, again, to continuations. Answer-type
modifying continuations can be packaged as an indexed monad. This was, in fact,
one of the original motivations for the introduction of indexed monads
\cite{indexed-monads}:

\begin{lstlisting}
newtype Cont r r' a
  = Cont { runCont :: (a -> r) -> r' }

instance IxMonad Cont where
  return x = Cont \k -> k x
  (Cont c) >>= f = Cont \k ->
    c \x -> runCont (f x) k
\end{lstlisting}

To connect this to what we've been doing so far, the type \hs{Tr r r'} of
\cref{sec:string-transformers} is isomorphic to a type of Kleisli arrows:
\hs{Tr r r' ≈ String -> Cont r r' String}.

\paragraph{Abstracting the Stack} In order to implement leads, like in
\cref{sec:leads}, we will need some way to access and modify the stack. Indexed
monads, by themselves, don't give us any access to the stack. We will need more
primitives. Let continuations guide us once again. The quintessential primitive of
\hs{Cont} is \hs{shift} \cite{abstracting-control}:
\begin{lstlisting}
shift :: ((a->r)->Cont k r' k) -> Cont r r' a
shift f = Cont \k -> runCont (f k) id
\end{lstlisting}
And, indeed, \hs{shift} lets us manipulate the stack. For instance, we can
implement a function to return the top of the stack:
\begin{lstlisting}
pop :: Cont r (a -> r) a
pop = shift \k -> return (\a -> k a)
\end{lstlisting}

However, \hs{Fwd} can't implement \hs{pop}, hence \hs{Fwd} can't implement
\hs{shift}. Indeed, in \hs{Fwd} the stack type is phantom. There isn't any stack
to return a value from. That is, we can't have stack-manipulating functions have
return values. Specializing \hs{shift} to the unit type gives us just that, and
forms the next step of our abstraction:
\begin{lstlisting}
class Stacked m where
  shift_ :: (r -> m k r' k) -> m r r' ()

instance Stacked Cont where
  shift_ f = shift \k -> f (k ())
\end{lstlisting}

The \hs{Stacked} type class lets us implement all sorts of
stack-manipulating operations. Note how, crucially, we can only
implement \hs{pop_}, a variant of \hs{pop} which simply discards the
value from the stack.
\begin{lstlisting}
push :: (IxMonad m, Stacked m)
     => a -> m (a -> r) r ()
push a = shift_ \k -> return (k a)

pop_ :: (IxMonad m, Stacked m) => m r (a->r) ()
pop_ = shift_ \k -> return (\_a -> k)

stack :: (IxMonad m, Stacked m)
      => (r -> r') -> m r r' ()
stack f = shift_ \k -> return (f k)
\end{lstlisting}

The \hs{stack} function, in particular, is a very flexible tool. It's
worth paying attention to the fact that because of how the answer types \hs{r} and \hs{r'} are
structured, a function \hs{r -> r'} really maps the stack \hs{r'} to
the stack \hs{r}. For instance in
\begin{lstlisting}
curryStack :: Cont ((a, b)->r) (a->b->r) ()
curryStack = stack \k a b -> k (a, b)
\end{lstlisting}
the type \hs{((a, b) -> r) -> (a -> b -> r)} is really the type
\hs{a -> b -> (a, b)} in CPS.

Together with, \emph{mutatis mutandis}, the interface of
\citet{combinator-parsers}, we now have our complete abstraction:

\begin{lstlisting}
class (Stacked m, IxMonad m) => Descr m where
  satisfy ::
     (Char -> Bool) -> m r (Char -> r) Char
\end{lstlisting}

\paragraph{Effects}
We have a complete interface of monadic format descriptors. But we don't
actually know how to print yet. In \cref{sec:tr} we used the type
\begin{lstlisting}
String -> Cont r r' String
\end{lstlisting}
as the type of printers. But, of course, this isn't a monad. We could try to fix up this type until we have a monad. But, since
monads are all about effects, it's better to think of printing as an effect that
we will add to our continuation monad.

It's tempting to try to encode effects with the usual continuation monad
transformer:
\begin{lstlisting}
newtype ContT m r r' a
  = ContT { runContT :: (a -> m r) -> m r' }

instance IxMonad (ContT m) where
  return x = ContT \k -> k x
  (ContT c) >>= f = ContT \k ->
    c (\x -> runContT (f x) k)
\end{lstlisting}
But \hs{ContT m} isn't an instance of \hs{Stacked}. In the below,
\begin{lstlisting}
shift_ :: Monad m => (r -> ContT m k r' k)
       -> ContT m r r' ()
shift_ f = ContT \k -> runCont (f _) return
\end{lstlisting}
we need a value of type \hs{r} to fill the hole, but we cannot extract one
from \hs{k :: () -> m r}. Furthermore, we cannot change \hs{shift_} to take a monadic
continuation either,
\begin{lstlisting}
badShift_ :: Monad m => (m r -> ContT m k r' k)
          -> ContT m r r' ()
badShift_ f = ContT \k ->
  runCont (f (k ())) return
\end{lstlisting}
as then we cannot, for instance, implement \hs{push} anymore. Perhaps more
concretely, how could we print a character with \hs{ContT}? In the below,

\begin{lstlisting}
satisfy :: ContT ((String,)) (Char -> r) r Char
satisfy _ = ContT \(k :: Char->(String, r)) ->
  (_ :: (String, Char -> r))
\end{lstlisting}
how can we fill the hole? Per the type, it must be of the form
\hs{(printed,\c -> _)}. That is, whatever is printed is
independent of \hs{c}. This isn't what we want. We would
really want the hole to be of type \hs{Char -> (String, r)}. But
this isn't a type that the monad transformer style can give us.

Let's look for our solution somewhere else: there's another, if less
travelled, way of complementing continuations with effects: using
comonads \cite{monads-from-comonads}. And this one enables us to implement
\hs{Stacked}.

\begin{lstlisting}
class Comonad w where
  extract :: w a -> a
  extend :: (w a -> b) -> w a -> w b

newtype ContW w r r' a
  = ContW { runContW :: w (a -> r) -> r' }

instance Comonad w => IxMonad (ContW w) where
  return x = ContW \wk -> extract wk x
  (ContW a) >>= f = ContW \wk ->
      a (extend k' wk)
    where k' wk x = runContW (f x) wk

instance Comonad w => Stacked (ContW w) where
  shift_ f = ContW \wk ->
    -- Continuation and its comonadic context
    -- are split, but both are passed to f.
    runContW (f (extract wk ()))
             (const id <$> wk)

yield :: (Comonad w)
       => (w r -> r) -> ContW w r r ()
yield eff = ContW \wk ->
  (eff ((\k -> k ()) <$> wk))
\end{lstlisting}
To recover the type \hs{Tr} of \cref{sec:tr}, we can use the \hs{Store} comonad:
\begin{lstlisting}
newtype Store s a = Store (s -> a, s)
\end{lstlisting}
Checking that \hs{ContW (Store String)} is indeed isomorphic to \hs{Tr} is left as an exercise to the reader.

However, in order to demonstrate the flexibility of this approach, and since
printing doesn't need the full power of state-passing, we can turn instead to
the \hs{Traced} comonad:
\begin{lstlisting}
newtype Traced m a
  = Traced { runTraced :: m -> a }
\end{lstlisting}
\begin{lstlisting}
instance Monoid m => Comonad (Traced m) where
  extract (Traced a) = a mempty
  extend f (Traced a) = Traced \m ->
    f (Traced \m' -> a (m <> m'))
\end{lstlisting}
Note that \hs{forall r.esc ContW (Traced m) r r a} is isomorphic to
\hs{(m, a)}. That is to say, \hs{ContW} is more precise than \hs{ContT}: \hs{forall r.esc ContT (m,) r r a} contains more computations than
\hs{(m, a)}. This is related to the observation that \hs{ContT m} is a
monad even if \hs{m} itself isn't a monad.
Key for us is the fact that the following two types are isomorphic,
\begin{lstlisting}
ContW w (Char -> r) r' a
Char -> ContW w r r' a
\end{lstlisting}
which is precisely what we were calling for.
This enables us to implement the \hs{satisfy} function:

\begin{lstlisting}
class Comonad w => ComTraced s w where
  trace :: s -> w a -> a

instance Monoid s =>
    ComTraced s (Traced s) where
  trace x (Traced f) = f x

instance ComTraced String w =>
    Descr (ContW w) where
  satisfy _ = pop >>= \c ->
    yield (trace [c]) *> return c
\end{lstlisting}

We can opt for
different pretty-printing libraries by varying the monoid \hs{s}. We use \hs{String} in our examples for the
sake of simplicity, but a pretty-printing library in the style of
\cite{prettier-printer} would use some \hs{Doc} type instead. Of course, there
is no conceptual difficulty in printing with an arbitrary pretty-printing
library with \hs{K7}-like combinators (reusing a parser library, on the other
hand, doesn't seem possible).

And with this, we have our monadic format descriptors. Let's revisit
Example~\ref{ex:scanf-printf} with stacked monads. Notice that because monads
don't have multiple return values, \hs{sscanf} isn't as flexible as in
\cref{sec:tr}: we need to choose the type in which the result is packaged in
\hs{spec} instead of being able to wait until the call to \hs{sscanf}. This is a
mild instance of tuple troubles which is inherent to monads (and applicatives).
It's usually fine because, contrary to the example of \cref{sec:analysis}, it
doesn't tend to create nested tuples. The full code for this section can be
found in \cref{sec:full-stacked-monad-linear}.

\begin{example}
We construct our type of format descriptors by pairing print and parse
format descriptors:
\begin{lstlisting}
data (pr :*: pa) r r' a = pr r r' a :*: pa r r' a

type D = ContW (Traced String) :*: Fwd Pa
\end{lstlisting}
On the parser side, a value of type \hs{Fwd Pa r r' a} is an
action of an indexed monad, over a monad \hs{Pa}.
We can take \hs{forall r.esc Tr r (a -> r)} as
\hs{Pa}, or just as simply:
\begin{lstlisting}
newtype Pa a = Pa (String -> (a, String))
\end{lstlisting}
\hs{sprintf}, \hs{sscanf} are given as follows:
\begin{lstlisting}
sprintf :: D String r a -> r
sprintf (ContW pr :*: _) = pr (Traced \s _ -> s)

sscanf :: D r r' a -> String -> a
sscanf (_ :*: Fwd (Pa pa)) s = fst (pa s)
\end{lstlisting}
Our \hs{spec} is as in Example~\ref{ex:scanf-printf}, but this time it
is not polyvariadic --- we have to map a tuple constructor over it:
\begin{lstlisting}
spec = (,,) <$>
  digit <* lit "-th character after "
  <*> char <* lit " is " <*> char
\end{lstlisting}
We can use \hs{spec} for both printing and parsing:
\begin{lstlisting}
>>> sprintf spec 5 'a' 'f'
"5-th character after a is f"

>>> sscanf spec "5-th character after a is f"
Just (5, 'a', f')
\end{lstlisting}
The inferred type of \hs{spec} is,
\begin{lstlisting}
spec :: (Descr m)
     => m r (Int -> Char -> Char -> r)
            (Int, Char, Char)
\end{lstlisting}
\end{example}

\subsection{Returning to Partiality}
\label{sec:indexed-partiality}

As a final part of our exploration, let us now handle failure and choice in the
monadic context. For regular monads, this is the role of the \hs{MonadPlus}
class. In our setting, we add an extra \hs{Monoid}
superclass as in \cref{sec:c-tr}:
\begin{lstlisting}
class (Stacked m, IxMonad m
      , forall r r' a!.! Monoid (m r r' a))
       => Descr m where
  satisfy ::
    (Char -> Bool) -> m r (Char -> r) Char
\end{lstlisting}

We might hope that we can use the comonad \hs{w} to add partiality as
an effect. However, this seems to be a dead end. Indeed, there is no comonad
\hs{W} such that \hs{ContT W r r} is isomorphic to \hs{Maybe} (or to
\hs{List} for that matter).

Therefore, like in \cref{sec:c-tr}, we use two
continuations. Remarkably, the implementation of the indexed monad
instance is exactly the same as that for \hs{ContW}:
\begin{lstlisting}
newtype Cont2W w r r' a = Cont2W
  { runCont2W :: w (a -> r -> r) -> r' -> r' }

instance Comonad w => IxMonad (Cont2W w) where
  return x = Cont2W \wk -> extract wk x
  Cont2W a >>= f = Cont2W \wk ->
      a (extend k' wk)
    where
      k' wk x = runCont2W (f x) wk
instance Comonad w =>
    Monoid (Cont2W w r r' a) where
  mempty = Cont2W \_ fl -> fl
  (Cont2W a) <> (Cont2W b) = Cont2W \wk fl ->
    a wk (b wk fl)
\end{lstlisting}
We will still use \hs{Fwd} and \hs{(:*:)}, as they preserve monoid. Their
instances, and the code supporting this section can be found in
\cref{sec:full-stacked-monad-plus}.

Iterating the CPS transform gives rise to the so-called CPS
hierarchy~\cite{abstracting-control}, and \hs{Cont2W} is the (comonad-enriched)
second iteration. There is, however, a bit of a design choice here. Indeed,
\hs{Cont2W} isn't the only type that we can give to said second iteration. For
instance:
\begin{lstlisting}
type C2 h r r' = (a -> h -> r) -> h -> r'
\end{lstlisting}
We could also give a type with four type parameters which is more general than
both, but it's also rather hard to tame into an abstraction.

The practical implication of this choice is that \hs{C2} doesn't support a
choice monoid at every type (indeed \hs{C2} doesn't even support \hs{mempty} at
every type). On the one hand, \hs{C2} implements the same \hs{shift} as
\cref{sec:stacked-monads}\footnote{In fact, \hs{C2 h} can be defined as \hs{ContW (StoreT h)}.}.
On the other hand, \hs{Cont2} does support choice at every type, but requires a
different type for \hs{shift}. So we need to
modify the \hs{Stacked} type class to account for the particular type of its continuations. \hs{Fwd} and \hs{(:*:)} are unaffected.
\begin{lstlisting}
class IxMonad m => Stacked m where
  shift_ :: ((r -> r) -> r' -> m k r' k)
         -> m r r' ()

instance Comonad w => Stacked (Cont2W w) where
  shift_ f = Cont2W \wk k' ->
    runCont2W (f (extract wk ()) k')
              ((\_k -> \x _ -> x) <$> wk) k'

push :: Stacked m => a -> m (a -> r) r ()
push x = shift_ \k k' ->
  return (k (\_ -> k') x)

pop_ :: Stacked m => m r (a -> r) ()
pop_ = shift_ \k k' -> return (\a -> k (k' a))

stack :: Stacked m => (r' -> r -> r')
      -> (r' -> r) -> m r r'
stack f u =
  shift_ \k k' -> return (f k' (k (u k')))
\end{lstlisting}
We can see two things happening in the type of \hs{stack}. First, the
function \hs{f} takes an additional argument of type \hs{r'}. This
lets us declare a failure to \hs{stack} --- useful to implement leads.
Then, \hs{stack} takes an extra argument \hs{u}: it's a “stack
unrolling” function. The role of \hs{u} is to restore the stack to its
initial state in case of failures. To illustrate this, implement the
lead for the \hs{(:)} constructor explicitly:
\begin{lstlisting}
consL :: Stacked m
      => m (a -> [a] -> r) ([a] -> r)
           (a -> [a] -> [a])
consl = stack uncons unroll *> return (:)
  where
    uncons k' k (x : xs) = k x xs
    uncons k' k [] = k' []

    unroll k' x xs = k' (x:xs)
\end{lstlisting}
This extra bookkeeping seems to be the price to pay to have a monoid
at every \hs{Cont2W w r r' a}, to match the usual expressive power of
\hs{MonadPlus}.

Following the same recipe as \hs{consL}, and just like in
\cref{sec:prisms}, we can implement a lead for any prism:
\begin{lstlisting}
prismL :: Stacked m => Prism' s a
       -> m (a -> r) (s -> r) (a -> s)
prismL l = stack rev u *> return (review l)
  where
    rev k' k t = case preview l t of
      Nothing -> k' t
      Just x -> k x

    u k' a = k' (review l a)
\end{lstlisting}

Finally, to parse a string, we apply the format descriptor, throwing
away what remains of the string. The implementation of pretty printing is analogous to
\cref{sec:case-lambda}:
\begin{lstlisting}
type D2 = Cont2W (Traced String) :*: Fwd Pa

parse :: D2 r r' b -> String -> Maybe b
parse (_ :*: Fwd (Pa pa)) s = fst <$> pa s

pretty
  :: D2 (Maybe String) (a -> Maybe String) b
  -> a -> Maybe String
pretty (Cont2W pr :*: _) =
  pr (Traced \s _ _ -> Just s) (\_ -> Nothing)
\end{lstlisting}

And this is it. We now have the material to build the
$\lambda$-calculus example of \cref{sec:case-lambda}. We do so in
\cref{fig:ex-lambda-pup}. Notice the following differences.

Where descriptors as cassettes are glued together uniformly using
\hs{(.)}, descriptors as stacked monads rely on a variety of operators
to selectively drop the return value on the left or right-hand sides.
This is needed because we no longer have true nullary descriptors.
Here, \hs{lit} returns \hs{()}, when it previously returned nothing at
all.

A more instructive observation is that in the \hs{App} case we
write
\begin{lstlisting}
parens (appL <*> term <* sepSpace <*> term)
\end{lstlisting}
where in \cref{sec:case-lambda} we could write
\begin{lstlisting}
appL --> parens (term . sepSpace . term)
\end{lstlisting}
It is not possible to write the same with stacked monads:
\begin{lstlisting}
appL <*> parens (term <* sepSpace <*> term)
  -- Type error
\end{lstlisting}
We could work around that by tupling and untupling. This is another
instance of tuple troubles lurking right around the corner in the
stacked monad style. There is an as yet unresolved tension here,
between the convenience of polyvariadic descriptors and the expressive
power of context sensitivity afforded by monads.

\begin{lfigure*}
\begin{lstlisting}[multicols=2,xleftmargin=0pt]
parens p = lit "(" *> p <* lit ")"
sepSpace = lit " "
idnt = consL <*> letter <*> many alphaNum



term :: (Descr m) => m r (Term -> r) Term
term =
  varL <*> idnt <>
  absL <* lit "!$\lambda$!" <*> idnt <* lit "!.!" <*> term <>
  parens (appL <*> term <* sepSpace <*> term)
\end{lstlisting}
  \caption{$\lambda$-calculus with stacked monads}
  \label{fig:ex-lambda-pup}
\end{lfigure*}

\begin{lfigure*}
\begin{lstlisting}
term :: Biparser Term Term
term = (Var <$> idnt) `uponr` _Var <>
       (pure Abs <* lit "!$\lambda$!" <*> idnt `upon'` fst <* lit "!.!" <*> term `upon'` snd) `uponr` _Abs <>
       parens ((App <$> term `upon'` fst <* sepSpace <*> term `upon'` snd) `uponr` _App)
\end{lstlisting}

\begin{minipage}{0.5\linewidth}
\begin{lstlisting}
upon' :: PMP p => p b c -> (a -> b) -> p a c
upon' p f = p `upon` (Just . f)
\end{lstlisting}
\end{minipage}
\begin{minipage}{0.4\linewidth}
\begin{lstlisting}
uponr :: PMP p => p a b -> Prism' s a
uponr p pat = p `upon` review pat
\end{lstlisting}
\end{minipage}

  \caption{$\lambda$-calculus with partial monadic profunctors}
  \label{fig:pmp-biparsers}
\end{lfigure*}

\section{Assessment}

\citet{functional-unparsing} sought to illustrate the ``expressive
power of ML'' by implementing \hs{printf} without dependent types.
The technique is furthermore a demonstration that
\emph{continuations get you out of tuple troubles}. This is again
demonstrated with the construction
of \hs{scanf} by \citet{functional-ununparsing} along the same lines, or by \citet{pattern-combinators} delightfully applying
the same technique to first-class patterns for pattern-matching.

\citet{hinze-formatting} demonstrates a direct-style solution, in
a language richer than ML, with type classes and functional
dependencies to build a type-level function over types.
\citeauthor{hinze-formatting} relies on functor
composition, which isn't associative in Haskell.
Therefore, similar troubles to tuple troubles arise.
In particular, \hs{a <> b} may or may not type check depending
on how functor compositions were nested in the respective types
of \hs{a}, \hs{b}.\unsure{Though, couldn't any associativity troubles be
  normalized away in a type-level function? [Arnaud] possibly, but then all your
types will carry unevaluated type families in them, which doesn't sound like a
fun time.}

The original motivation of the arrows of \citet{arrows} was to capture
the essence of the parser combinators of \citet{combinator-parsers}.
But in the end, \citeauthor{arrows} uses only a weaker fragment
\cite{idioms-oblivious} of
arrows that \citet{applicative} identify as \hs{Alternative}. In
general though, programming with nested tuples is at the core of the
arrow style \cite{programming-arrows}, to the point that
\citet{proc-notation} proposes dedicated syntactic sugar to hide the
tupling. \citet{biarrows} uses this syntactic sugar to implement
combinators as invertibility-preserving pairs, much like our format
descriptors. More recent developments also adopt this style
\cite{biparser-exact-print}. \citet{partial-isomorphism} achieve much
the same in applicative style, but again perform monoidal aggregation
via nested pairs.

To the best of our knowledge, the \hs{K7} descriptors of \cref{sec:tr,sec:c-tr} is
the only set of combinators that avoids tuple troubles entirely,
but like the above cited works, at the price of
only parsing context-free grammars. \citet{monadic-profunctors} achieve greater expressivity, still with nested pairs.
Even the stacked monads of
\cref{sec:pup} exhibit some tuple troubles. More research would be
needed to find a tuple-trouble-free context-sensitive set of format
descriptors. In particular, it feels like what we are grasping at
there is a notion of multiple-result monads yet to be developed, akin to
the multiple-result functions of \cref{sec:tr}. In essence, it would
appear that the ``potential equivalence'' between the CPS hierarchy
and monads that \citet{abstracting-control} were musing about has not,
35 years later, been fully resolved.

\Citet{monadic-profunctors} introduce a notion of \emph{partial
monadic profunctor} (PMP) to give an interface to format descriptors:

\begin{lstlisting}
class (forall a!.! MonadPlus (p a)) => PMP p where
  upon :: p a b -> (a' -> Maybe a) -> p a' b
\end{lstlisting}

The intent is that a \hs{p a b} can print an \hs{a} and parse into a
\hs{b}. This design was the main inspiration behind the stacked monads
of \cref{sec:stacked-monads}, which can be summarised as applying a
continuation-based stack to the contravariant argument to overcome the
inherent tuple troubles of the \hs{Comonoidal} type class (as analysed
in \cref{sec:analysis}). Format descriptors implemented with stacked
monad are significantly less verbose. Compare for instance the syntax
of the $\lambda$-calculus with PMPs in \cref{fig:pmp-biparsers}, with
that of \cref{fig:ex-lambda-pup}.

Note that \hs{PMP}s are, indeed, comonoidal as another equivalent type for
\hs{comult} is
\begin{lstlisting}
comult'' :: Comonoidal f => f r -> f r -> f r
\end{lstlisting}
There's no tuple in this interface, but, as is apparent in
\cref{fig:pmp-biparsers}, \hs{comult''} only replaces automatic nested tupling
with the need to manually re-adjust the base type everywhere; \emph{e.g.}
pairing each descriptor with a call to \hs{upon}.

It's worth noting, however, that PMPs can implement failure and choice
abstractly using a monad transformer (\hs{MaybeT}), whereas this
doesn't seem possible for stacked monads as discussed in
\cref{sec:indexed-partiality}.

\paragraph{Further Considerations}
The developments presented in this paper ignore several concerns that
are important in practice, like good error messages, precise location
of the cause of these errors, potential space leaks due
to unrestricted backtracking \cite{parsec}, threading user state, and
controlling the printer's layout.
All of the above could in principle be worked into our framework. But
a more practical and maintainable approach is to reuse off-the-shelf,
mature parser and pretty printing combinator libraries. This is
demonstrated in the libraries accompanying this paper\footnote{
  \url{https://github.com/mboes/cassette/} and \url{https://github.com/tweag/pup}
}. We don't know, however, how to reuse parsing
libraries in the \hs{K7} style. Once again there's a tension left to
be resolved between \hs{K7}'s tuple trouble freedom and practical
needs.

Another venue for further investigation is the relation between the
comonads of \cref{sec:pup} and delimited
control. Indeed, \citet{typed-functional-unparsing} demonstrates how
\citeauthor{functional-unparsing}'s \hs{printf} can be reinterpreted
without explicit continuation passing in a (typed,
answer-type-modifying) language with \hs{shift} and \hs{reset}. Yet,
if passing continuations explicitly lets us wrap the continuation in a
comonad, how could we do the same in a language with delimited control
instead? In this way, \citeauthor{typed-functional-unparsing}'s solution
could extend, generically, to more applications than just
printing.

Beyond the applications to format descriptors, we might imagine that stacked
functors or categories can find other uses in functional programming.
\Citet{monadic-profunctors} also apply PMPs to defining multiple-focus lenses,
\citet{reflect-random} proposes several applications to random generation.
Presumably, wherever PMPs are useful, stacked monads should apply too. But
beyond even bidirectional programming, any interface using contravariant
functors or profunctors as an abstraction is liable to have some amount of tuple
trouble. We argue that even though they aren't as grounded in mathematical
practice, stacked functors and categories may be more natural in functional
programming languages. Therefore, we look forward to seeing whether such interfaces
can be revisited with stacks to good effect.

\bibliographystyle{ACM-Reference-Format}
\bibliography{bibliography}{}

\appendix

\section{Full Linear Scanning Example with Stacked Monads}
\label{sec:full-stacked-monad-linear}

\begin{lstlisting}
class (Functor w) => Comonad w where
  extract :: w a -> a
  extend :: (w a -> b) -> w a -> w b

newtype Traced m a = Traced {runTraced :: m -> a}
  deriving (Functor)

instance (Monoid m) => Comonad (Traced m) where
  extract (Traced a) = a mempty
  extend f (Traced a) = Traced \m ->
    f (Traced \m' -> a (m <> m'))

class (Comonad w) => ComTraced m w where
  trace :: m -> w a -> a

instance (Monoid m) => ComTraced m (Traced m) where
  trace x (Traced f) = f x

newtype Pa a
  = Pa {runPa :: String -> (a, String)}
  deriving (Functor)

instance Monad Pa where
  -- return a = Pa \s -> Just (a, s)
  (Pa p) >>= f = Pa \s ->
    let (a, s') = p s
     in runPa (f a) s'

instance Applicative Pa where
  pure a = Pa \s -> (a, s)
  (<*>) = ap

data (f :*: g) r r' a
  = (:*:) {ifst :: (f r r' a), isnd :: (g r r' a)}

instance
  (IxMonad f, IxMonad g) =>
  IxMonad (f :*: g)
  where
  return x = return x :*: return x
  ~(l :*: r) >>= f =
    (l >>= (ifst . f)) :*: (r >>= (isnd . f))

newtype Fwd m r r' a = Fwd {unFwd :: m a}

instance (Monad m) => IxMonad (Fwd m) where
  return x = Fwd (return x)
  (Fwd a) >>= f = Fwd (a >>= (unFwd . f))

class Stacked m where
  shift_ :: (r -> m k r' k) -> m r r' ()

push :: (IxMonad m, Stacked m)
     => a -> m (a -> r) r ()
push a = shift_ \k -> return (k a)

pop_ :: (IxMonad m, Stacked m) => m r (a -> r) ()
pop_ = shift_ \k -> return (\_a -> k)

stack :: (IxMonad m, Stacked m)
      => (r -> r') -> m r r' ()
stack f = shift_ \k -> return (f k)

newtype Cont r r' a
  = Cont {runCont :: (a -> r) -> r'}

instance IxMonad Cont where
  return x = Cont \k -> k x
  (Cont c) >>= f = Cont \k ->
    c (\x -> runCont (f x) k)

instance Stacked Cont where
  shift_ f = shift \k -> f (k ())

shift :: ((a -> r) -> Cont k r' k) -> Cont r r' a
shift f = Cont \k -> runCont (f k) id

pop :: Cont r (a -> r) a
pop = shift \k -> return (\a -> k a)

instance
  (Stacked f, Stacked g) =>
  Stacked (f :*: g)
  where
  shift_ f = (shift_ (ifst . f) :*: shift_ (isnd . f))

instance (Monad m) => Stacked (Fwd m) where
  shift_ _f = Fwd (return ())

class (Stacked m, IxMonad m) => Descr m where
  satisfy :: (Char -> Bool) -> m r (Char -> r) Char

instance (Descr f, Descr g) => Descr (f :*: g) where
  satisfy p = satisfy p :*: satisfy p

instance Descr (Fwd Pa) where
  satisfy p = Fwd (Pa go)
    where
      go (c : s) | p c = (c, s)

newtype ContT m r r' a
  = ContT {runContT :: (a -> m r) -> m r'}

instance IxMonad (ContT m) where
  return x = ContT \k -> k x
  (ContT c) >>= f = ContT \k ->
    c \x -> runContT (f x) k

lift :: (Monad m) => m a -> ContT m r r a
lift act = ContT \k -> act >>= k

newtype ContW w r r' a
  = ContW {runContW :: w (a -> r) -> r'}

shiftw :: (Comonad w)
       => ((a -> r) -> ContW w k r' k)
       -> ContW w r r' a
shiftw f = ContW \wk ->
  runContW (f (extract wk)) (const id <$> wk)

popw :: (Comonad w) => ContW w r (a -> r) a
popw = shiftw \k -> return (\a -> k a)

instance (Comonad w) => IxMonad (ContW w) where
  return x = ContW \wk -> extract wk x
  (ContW a) >>= f = ContW \wk -> a (extend k' wk)
    where
      k' wk x = runContW (f x) wk

instance (Comonad w) => Stacked (ContW w) where
  shift_ f = ContW \wk ->
    -- Notice how the continuation and its
    -- comonadic context are split, but both
    -- are passed to f.
    runContW (f (extract wk ())) (const id <$> wk)

yield :: (Comonad w)
      => (w r -> r) -> ContW w r r ()
yield eff = ContW \wk ->
  (eff ((\k -> k ()) <$> wk))

instance (ComTraced String w)
  => Descr (ContW w) where
  satisfy _ =
    popw >>= \c ->
      yield (trace [c]) *> return c

type D = ContW (Traced String) :*: Fwd Pa

lit :: (Descr m) => String -> m r r ()
lit [] = return ()
lit (c : cs) =
  push c *> satisfy (== c) >>= \_ -> lit cs

char :: (Descr m) => m r (Char -> r) Char
char = satisfy (const True)

digit :: (Descr m) => m r (Int -> r) Int
digit = return (\c -> read [c]) <*
        stack (\k -> k . head . show) <*>
        satisfy isDigit

spec :: (Descr m)
     => m r (Int -> Char -> Char -> r)
            (Int, Char, Char)
spec = (,,) <$>
  digit <* lit "-th character after "
  <*> char <* lit " is " <*> char

-- | Ex:
-- >>> sprintf spec 5 'a' 'f'
-- "5-th character after a is f"
sprintf :: D String r a -> r
sprintf (ContW pr :*: _) = pr (Traced (\s _ -> s))

-- | Ex:
-- >>> sscanf spec "5-th character after a is f"
-- Just (5, 'a', f')
sscanf :: D r r' a -> String -> a
sscanf (_ :*: Fwd (Pa pa)) s = fst (pa s)
\end{lstlisting}

\section{Full $\lambda$-Calculus Example with Stacked Monads}
\label{sec:full-stacked-monad-plus}

\begin{lstlisting}
newtype Pa a
  = Pa {runPa :: String -> Maybe (a, String)}
  deriving (Functor)

instance Monad Pa where
  -- return a = Pa \s -> Just (a, s)
  (Pa p) >>= f = Pa \s -> do
    ~(a, s') <- p s
    runPa (f a) s'

instance Applicative Pa where
  pure a = Pa \s -> Just (a, s)
  (<*>) = ap

instance MonadPlus Pa

instance Alternative Pa where
  empty = Pa \_ -> empty
  (Pa pa1) <|> (Pa pa2) = Pa \s -> pa1 s <|> pa2 s

class (Stacked m, IxMonad m
      , forall r r' a!.! Monoid (m r r' a))
      => Descr m where
  satisfy :: (Char -> Bool) -> m r (Char -> r) Char

instance (Descr f, Descr g) => Descr (f :*: g) where
  satisfy p = satisfy p :*: satisfy p

instance Descr (Fwd Pa) where
  satisfy p = Fwd (Pa go)
    where
      go (c : s) | p c = Just (c, s)
      go _ = Nothing

newtype Cont2W w r r' a = Cont2W
  {runCont2W :: w (a -> r -> r) -> r' -> r'}

instance (Comonad w) => IxMonad (Cont2W w) where
  return x = Cont2W \wk -> extract wk x
  Cont2W a >>= f = Cont2W \wk -> a (extend k' wk)
    where
      k' wk x = runCont2W (f x) wk

instance (Comonad w)
  => Monoid (Cont2W w r r' a) where
  mempty = Cont2W \_ fl -> fl
  (Cont2W a) <> (Cont2W b) = Cont2W \wk fl ->
    a wk (b wk fl)

shiftw :: (Comonad w)
       => ((a -> r -> r) -> r' -> Cont2W w k r' k)
       -> Cont2W w r r' a
shiftw f = Cont2W \wk k' ->
  runCont2W (f (extract wk) k')
            (const (\k _ -> k) (<$>) wk) k'

pop :: (Comonad w) => Cont2W w r (a -> r) a
pop = shiftw \k k' -> return (\a -> k a (k' a))

instance (ComTraced String w)
  => Descr (Cont2W w) where
  satisfy _ =
    pop >>= \c ->
      yield (trace [c]) *> return c

yield :: (Comonad w)
      => (w r -> r) -> Cont2W w r r ()
yield eff = Cont2W \wk k' ->
  eff ((\k -> k () k') <$> wk)

instance (MonadPlus m)
  => Monoid (Fwd m r r' a) where
  mempty = Fwd empty
  (Fwd a) <> (Fwd b) = Fwd (a <|> b)

instance MonadPlus Prs

instance Alternative Prs where
  empty = Prs \_ -> empty
  (Prs pa1) <|> (Prs pa2) = Prs \s ->
    pa1 s <|> pa2 s

instance
  (Monoid (f r r' a), Monoid (g r r' a)) =>
  Monoid ((f :*: g) r r' a)
  where
  mempty = mempty <> mempty
  ~(fl :*: fr) <> ~(gl :*: gr) =
    (fl <> gl) :*: (fr <> gr)

class (IxMonad m) => Stacked m where
  shift_ ::
    ((r -> r) -> r' -> m k r' k) ->
    m r r' ()

instance (Comonad w) => Stacked (Cont2W w) where
  shift_ f = Cont2W \wk k' ->
    runCont2W
      (f (extract wk ()) k')
      ((\_k -> \x _ -> x) <$> wk)
      k'

push :: (Stacked m) => a -> m (a -> r) r ()
push x = shift_ \k k' -> return (k (\_ -> k') x)

pop_ :: (Stacked m) => m r (a -> r) ()
pop_ = shift_ \k k' -> return (\a -> k (k' a))

stack ::
  (Stacked m) =>
  (r' -> r -> r') ->
  (r' -> r) ->
  m r r' ()
stack f u =
  shift_ \k k' -> return (f k' (k (u k')))

instance (Monad m) => Stacked (Fwd m) where
  shift_ _f = Fwd (return ())

instance (Stacked f, Stacked g)
  => Stacked (f :*: g) where
  shift_ f =
    (shift_ (\k k' -> ifst (f k k'))) :*:
    (shift_ (\k k' -> isnd (f k k')))

consL ::
  (Stacked m) =>
  m (a -> [a] -> r)
    ([a] -> r)
    (a -> [a] -> [a])
consL = stack uncons unroll *> return (:)
  where
    uncons k' k (x : xs) = k x xs
    uncons k' k [] = k' []

    unroll k' x xs = k' (x : xs)

data Prism' s a
  = Prism' { review :: a -> s
           , preview :: s -> Maybe a }

prismL ::
  (Stacked m) =>
  Prism' s a ->
  m (a -> r) (s -> r) (a -> s)
prismL l = stack rev u *> return (review l)
  where
    rev k' k t = case preview l t of
      Nothing -> k' t
      Just x -> k x

    u k' a = k' (review l a)

some :: (Descr m)
     => (forall r!.! m r (a -> r) a)
     -> m r' ([a] -> r') [a]
some p = consL <*> p <*> many p

many :: (Descr m)
     => (forall r. m r (a -> r) a)
     -> m r' ([a] -> r') [a]
many p = some p <> (pop_ *> return [])

lit :: (Descr m) => String -> m r r ()
lit [] = return ()
lit (c : cs) =
  push c *> $satisfy (== c) >>= \_ -> lit cs

letter =
  satisfy (\c -> isLetter c && isAscii c)

alphaNum =
  satisfy (\c -> isAlphaNum c && isAscii c)

parens :: D2 r r' a -> D2 r r' a
parens p = lit "(" *> p <* lit ")"

sepSpace = lit " "

idnt = consL <*> letter <*> many alphaNum

type Idnt = String

data Term =
  Var Idnt | Abs Idnt Term | App Term Term
  deriving (Show)

type D2 = Cont2W (Traced String) :*: Fwd Pa

term :: D2 r (Term -> r) Term
term =
  varL <*> idnt <>
  absL <* lit "!$\lambda$!" <*> idnt <* lit "!.!" <*> term <>
  parens (appL <*> term <* sepSpace <*> term)

varL :: (Descr m)
     => m (Idnt -> r) (Term -> r) (Idnt -> Term)
varL = prismL _Var
  where
    _Var = Prism'
      { review = Var
      , preview = \case
          Var x -> Just x
          _ -> Nothing }

absL :: (Descr m)
     => m (Idnt -> Term -> r)
          (Term -> r)
          (Idnt -> Term -> Term)
absL =
  stack (\k' k -> \case Abs x u -> k x u; t -> k' t)
        (\k x u -> k (Abs x u))
  *> return Abs

appL :: (Descr m)
     => m (Term -> Term -> r)
          (Term -> r)
          (Term -> Term -> Term)
appL =
  stack (\k' k -> \case App u v -> k u v; t -> k' t)
        (\k u v -> k (App u v))
  *> return App

-- |
-- >>> parse term "!$\lambda$!x.(x x)"
-- Just (Abs "x" (App (Var "x") (Var "x")))
parse :: D2 r r' b -> String -> Maybe b
parse (_ :*: Fwd (Pa pa)) s = fst <$> pa s

-- |
-- >>> (parse term "!$\lambda$!xx.(x x)") >>= pretty term
-- Just "!$\lambda$!x.(x x)"
pretty :: D2 (Maybe String) (a -> Maybe String) b
       -> a -> Maybe String
pretty (Cont2W pr :*: _) =
  pr (Traced (\s _ _ -> Just s)) (\_ -> Nothing)
\end{lstlisting}

\end{document}